\documentclass[10pt,aps,prb,twocolumn,superscriptaddress,floatfix,longbibliography]{revtex4-2}
\usepackage{placeins}
\usepackage{mathtools}
\usepackage{physics}
\usepackage{bbm}
\usepackage{bm}
\usepackage{amsbsy}
\usepackage{amsthm}
\usepackage{amssymb}
\usepackage{amsfonts}
\usepackage{amsmath}
\usepackage{soul}
\setstcolor{red}
\usepackage{dsfont} 
\usepackage{graphicx} 
\usepackage{hhline}
\usepackage{epsfig}
\usepackage{epstopdf}
\usepackage{dsfont}
\usepackage{xcolor}
\usepackage{color}
\usepackage{verbatim}
\usepackage{nomencl}
\usepackage[normalem]{ulem}
\usepackage{multirow}
\usepackage{makecell}
\usepackage{url}
\usepackage{tikz}
\usetikzlibrary{arrows.meta, decorations.pathreplacing}
\usepackage[colorlinks=true,linkcolor=blue,citecolor=blue,urlcolor=blue]{hyperref}

\newcommand{\blue}[1]{\textcolor{blue}{#1}}

\usepackage{placeins}
\begin{document}

\title{Topological properties and Majorana Multiplicity in Zigzag Kitaev Chain}
\author{Rajiv Kumar} 
\email{rajivkumar.rs.phy22@iitbhu.ac.in}
\affiliation{Department of Physics, \blue{Indian Institute of Technology (Banaras Hindu University)}, Varanasi - 221005, India}
\author{Panch Ram}
\email{panchram.phy@itbhu.ac.in}
\affiliation{Department of Physics, \blue{Indian Institute of Technology (Banaras Hindu University)}, Varanasi - 221005, India}
\author{Levan Chotorlishvili}
\affiliation{Department of Physics and Medical Engineering, \blue{Rzeszow University of Technology,} 35-959 Rzeszow Poland}
\author{Sunil K. Mishra}
\email{sunilkm.app@iitbhu.ac.in}
\affiliation{Department of Physics, \blue{Indian Institute of Technology (Banaras Hindu University)}, Varanasi - 221005, India}

\begin{abstract}
We investigate the spectral and topological properties of a zigzag Kitaev chain constructed from two diagonally coupled one dimensional Kitaev chains with a zero and finite superconducting pairing phase difference. Using a Bogoliubov-de Gennes formulation, we analyze the energy spectrum, distribution of Majorana zero modes (MZMs), the quasi-particle dispersion, and the winding number, respectively. For a zero phase difference, the resulting energy spectrum shows topological phases with two, four MZMs, and trivial regions. The phases of gap closure determine the topological phase boundaries. In particular, for the phase difference between $\phi=\pi$, the degeneracy of MZMs is partially lifted, leading to modified topological phases compared to the case $\phi=0$. The topological and trivial phase boundaries are further confirmed by evaluating the quasi-particle dispersion and the topological invariant, namely the winding number. We show that the zigzag Kitaev chain contributes independently to the total winding number $\nu = 1$ and $2$, giving rise to distinct topological phases that support two and four MZMs. The $\nu = 0$ gives rise to a trivial region. The energy spectrum of systems corroborates the analytical phase boundaries and reveals characteristics associated with hybridization, enabling us to obtain the complete phase diagram of the zigzag model. Our results establish the zigzag Kitaev chain as a minimal platform for engineering MZM quantum computations, with potential applications in the study of topological phases and Majorana based qubit physics.
\end{abstract}
\maketitle
\section{Introduction}\label{Introduction}
The study of topological superconductivity, a new phase of matter, has opened new directions in condensed matter physics \cite{Wang2017, PhysRevLett.102.187001, PhysRevLett.107.217001} primarily due to the emergence of exotic quasi-particle excitations with non-Abelian statistics \cite{sarma2015majorana, flensberg2021engineered} and robust boundary signatures. These excitations correspond to Majorana zero modes (MZMs), quasi-particles that are their own antiparticles \cite{PhysRevLett.50.1427, PhysRevLett.121.076802}. The Kitaev chain \cite{Kitaev_2001, Alicea_2012} and vortex-bound states in a two-dimensional \(p_x + i p_y\) superconductor \cite{PhysRevB.61.10267, PhysRevB.73.014505, PhysRevB.88.180503} provide a conceptually transparent platform for studying MZMs.  The Kitaev chain can be effectively realized in a semiconductor-superconductor nanowire with a sizable Rashba spin-orbit interaction \cite{PhysRevLett.100.096407, PhysRevLett.104.040502, PhysRevLett.105.177002, PhysRevB.82.134521, PhysRevB.81.125318, PhysRevB.93.155402, van_Zanten_2020, Yang_2020} and planar Josephson junctions \cite{PhysRevX.7.021032, PhysRevLett.118.107701}. Recent experiments employing nanowires have reported signatures \cite{PhysRevLett.109.186802, Das2012, PhysRevB.109.035415, Dvir_2023}. The possibility of hosting MZMs has attracted considerable interest due to their fundamental importance and potential technological applications \cite{PhysRevX.11.011015, PhysRevResearch.5.023088, ten_Haaf_2024}, which enable topologically protected braiding operations and rendering them promising building blocks for topologically protected quantum computation \cite{RevModPhys.80.1083, PhysRevLett.134.096601, PhysRevLett.120.230405, PhysRevLett.101.010501, PhysRevLett.94.166802, KITAEV20032}. Individually, MZMs have also attracted considerable attention in recent years \cite{VanLoo2026, 7cqp-ws6c, PhysRevB.92.104514, RevModPhys.87.137, pzr7-cw4c,  d1bl-p49r, t264-dq2z, fxs7-wdqp, 2qxk-wbs6}. This interest stems not only from their potential applications in robust quantum technologies, but also from substantial experimental progress toward their detection \cite{PhysRevLett.100.096407, PhysRevB.81.125318, PhysRevLett.104.040502, PhysRevLett.105.177002, PhysRevB.82.134521, science.aaf3961, science.1222360}. Despite significant progress, establishing the topological behaviours of MZMs in semiconductor-superconductor nanowires remains a nontrivial task. Various challenges, including disorder~\cite{Sau_2012, PhysRevB.103.224505}, impurity~\cite{PhysRevB.110.214506, PhysRevApplied.16.054053}, and quasi-MZMs of near-zero energy, can generate signatures that closely resemble those expected from topological MZMs, thereby complicating their unambiguous identification. This motivates the search for alternative platforms capable of hosting MZMs in a more controlled manner \cite{PhysRevB.93.155402, PhysRevB.95.235305}. However, in a minimal Kitaev chain \cite{PhysRevB.109.035415, vanLoo_2026}, such states arise only in a finely tuned region in the parameters, commonly referred to as ``sweet spots''. Consequently, they lack intrinsic topological protection and are sensitive to parameter tuning; MZMs in minimal Kitaev chains are commonly termed poor man’s Majoranas \cite{PhysRevB.111.115419, wptk-lvc5}.
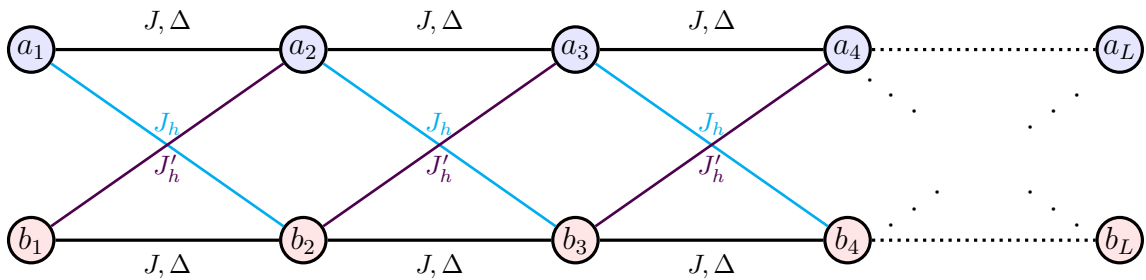
\begin{figure*}
\begin{tikzpicture}[scale=1.2,>=Stealth,
    site/.style={circle,draw=black,very thick,fill=#1!10,minimum size=6mm,inner sep=0pt},
    topcol/.style={site=blue},
    botcol/.style={site=red},
    link/.style={line width=1.2pt,black},
    linkdot/.style={line width=1.2pt,black,dotted},
    diagA/.style={font=\large,line width=1.0pt, draw=cyan!100!black},
    diagB/.style={font=\large,line width=1.0pt, draw=violet!60!black},
    lab/.style={font=\large, inner sep=1pt},
    diagAdot/.style={line width=1.0pt, draw=cyan!100!black, dotted},
    diagBdot/.style={line width=1.0pt, draw=violet!60!black, dotted},
]

\def\dx{3.0}
\def\dy{2.1}

\node[topcol] (T1) at (0,0) {\large{$a_1$}};
\node[topcol] (T2) at (\dx,0) {\large{$a_2$}};
\node[topcol] (T3) at (2*\dx,0) {\large{$a_3$}};
\node[topcol] (T4) at (3*\dx,0) {\large{$a_4$}};
\node[topcol] (T5) at (4*\dx,0) {\large{$a_L$}};

\node[botcol] (B1) at (0,-\dy) {\large{$b_1$}};
\node[botcol] (B2) at (\dx,-\dy) {\large{$b_2$}};
\node[botcol] (B3) at (2*\dx,-\dy) {\large{$b_3$}};
\node[botcol] (B4) at (3*\dx,-\dy) {\large{$b_4$}};
\node[botcol] (B5) at (4*\dx,-\dy) {\large{$b_L$}};

\draw[link] (T1) -- (T2) node[midway,above=2pt] {\normalsize{$J,\Delta$}};
\draw[link] (T2) -- (T3) node[midway,above=2pt] {\normalsize{$J,\Delta$}};
\draw[link] (T3) -- (T4) node[midway,above=2pt] {\normalsize{$J,\Delta$}};
\draw[linkdot] (T4) -- (T5) node[midway,above=2pt] {\normalsize{}};

\draw[link] (B1) -- (B2) node[midway,below=2pt] {\normalsize{$J,\Delta$}};
\draw[link] (B2) -- (B3) node[midway,below=2pt] {\normalsize{$J,\Delta$}};
\draw[link] (B3) -- (B4) node[midway,below=2pt] {\normalsize{$J,\Delta$}};
\draw[linkdot] (B4) -- (B5) node[midway,below=2pt] {\normalsize{}};

\draw[diagA] (T1) to[bend left=0] node[midway,above=-0.0pt,text=cyan!100!black] {\normalsize{$J_h$}} (B2);
\draw[diagA] (T2) to[bend left=0] node[midway,above=-0.0pt,text=cyan!100!black] {\normalsize{$J_h$}} (B3);
\draw[diagA] (T3) to[bend left=0] node[midway,above=-0.0pt,text=cyan!100!black] {\normalsize{$J_h$}} (B4);
\foreach \p in {0.01,0.1,0.2,0.7,0.8,0.9}
{\path (T4) -- (B5) node[pos=\p, below] {\large{$\cdot$}};}

\draw[diagB] (B1) to[bend left=0] node[midway,below=-0.0pt,text=violet!60!black] {\normalsize{$J_h'$}} (T2);
\draw[diagB] (B2) to[bend left=0] node[midway,below=-0.0pt,text=violet!60!black] {\normalsize{$J_h'$}} (T3);
\draw[diagB] (B3) to[bend left=0] node[midway,below=-0.0pt,text=violet!60!black] {\normalsize{$J_h'$}} (T4);
\foreach \p in {0.1,0.2,0.3,0.7,0.8,0.9}
{\path (B4) -- (T5) node[pos=\p, below] {\large{$\cdot$}};}
\end{tikzpicture}
\caption{Schematic illustration of the zigzag Kiatev chain: (top: $a_i$, bottom: $b_i$) of length $L$. In the calculation, we consider $L=40$ and a chain length $ N=2L=80$. Each chain features hopping amplitude $J$ and superconducting pairing strength $\Delta$. The cyan bonds ($J_h$) denote diagonal inter-chain couplings between sites $(a_i, b_{i+1})$ and the violet bonds ($J_h'$) connect $(b_i, a_{i+1})$. These diagonal couplings mediate hybridization between the two Kitaev chains.}
\label{two_kitaev_parallel_1}
\end{figure*}

Given the rapid developments in both theoretical proposals and experimental realizations, the Kitaev chain offers a feasible route for realizing MZMs, as their properties crucially determine the stability and localization of these excitations. Alongside experimental developments, theoretical model construction has played a central role in the systematic search for MZMs. Motivated by these considerations, a wide variety of Kitaev-type models have been proposed, including long-range Kitaev chains \cite{PhysRevB.95.195160, PhysRevLett.113.156402, PhysRevLett.119.110601, PhysRevB.104.075113, PhysRevB.94.125121}, dimerized Kitaev chains \cite{PhysRevB.90.014505, PhysRevB.96.205428, PhysRevB.101.184514, PhysRevB.100.205302}, and hybrid Kitaev chains \cite{7cqp-ws6c,n7bl-slgm, PhysRevB.99.094523}. Beyond providing realizations of MZMs, these systems host a rich variety of topological phases characterized by distinct bulk invariants and corresponding edge modes, making them versatile platforms for exploring edge physics. Extending beyond one dimensional systems, quasi one dimensional systems arise as an important intermediate class when moving toward higher dimensions, exhibiting a range of novel and enhanced features. Representative examples include coupled chains and ladder geometries, where recent studies have uncovered various topological phases, including the emergence of topological behaviour in Kitaev ladder systems. Recent efforts have further generalized the Kitaev model to incorporate ladder geometries and Creutz-ladder-type lattices \cite{ PhysRevB.107.075422, kxqp-57vj, PhysRevB.111.224203, tswn-kxhx, hfkz-jptt, PhysRevB.79.214435, PhysRevB.106.224306, PhysRevB.106.205111, PhysRevB.108.045415, PhysRevResearch.5.L012009, PhysRevB.106.125155, ml88-hbd5, f1m4-vkvq, rf8j-lf4s, kumar2026majorana_hybrid}.  These extensions have revealed a rich topological phase and enabled investigations of ladder coupling effects. Nevertheless, the topological characterization of more generalized ladder architectures, particularly those involving diagonal couplings, remains comparatively unexplored. Motivated by these developments, we extended Kitaev chains with diagonal couplings and investigated their spectral and topological properties. This geometry provides an additional tunable degree of freedom that reshapes the energy spectrum and enables the emergence of multiple MZM sectors, offering a natural platform for studying unconventional mechanisms of topological phases.

In this paper, we investigate an extension of the Kitaev chain, namely the zigzag Kitaev chain, which is composed of two Kitaev chains connected by asymmetric diagonal inter-chain couplings. We present a detailed theoretical investigation of the zigzag Kitaev chain, demonstrating that the system exhibits a rich topological phase diagram. The topological phases are characterized by using two complementary approaches: analysis of the energy spectrum to identify transition points and evaluation of the momentum-space topological invariant, i.e. the winding number. The complete agreement between these methods establishes a consistent bulk-boundary correspondence and enables a precise identification of topological phase transitions. The MZMs have attracted considerable interest for topological quantum computation. In particular, qubit architectures such as tetrons encode quantum information nonlocally in spatially separated MZMs, providing intrinsic protection against decoherence~\cite{qx36-4rv1}. In this context, the zigzag Kitaev chain systems offer a minimal and analytically tractable framework to investigate the emergence, control, and hybridization of multiple MZMs. The zigzag Kitaev chain studied here serves as a minimal lattice realization where diagonal inter-chain couplings and chemical potential provide tunable control over multiple MZMs and hybridization. This enables access to regimes supporting four MZMs, corresponding to a qubit subspace, as well as regimes where degeneracy is partially lifted due to hybridization. Our results, therefore, establish a direct connection between microscopic lattice models of topological superconductors and architectures relevant for Majorana based quantum quantum computing.

The paper is organized as follows. In Sec.~\ref{Model}, we investigate the zigzag Kitaev chain with its energy spectrum, bulk properties, dispersion relations, and winding number. Sec.~\ref{PD_ZKC} is examined for analysis of the corresponding phase diagram. Finally, Sec.~\ref{Conclusion} summarizes the main findings and provides an outlook for future work.
\section{Model}
\label{Model}
\subsection{Zigzag Kitaev chain}\label{TWO_ZKC}
We extend the Kitaev chain to a zigzag geometry composed of two parallel one-dimensional Kitaev chains coupled through diagonal inter-chain coupling terms in open boundary condition
(OBC). This model serves as a minimal platform for hosting multiple MZMs per edge and realizing multichannel topological phases. The Hamiltonian of the zigzag Kitaev chain is
\begin{eqnarray}
    H = H_1 + H_2 + H_{12},
    \label{H_ZKC1} 
\end{eqnarray}
where $H_r$ for $r\in (1,2)$ denotes the two Kitaev chains and $H_{12}$ represents the coupling between the chains. The first Kitaev chain Hamiltonian $H_1$ is described by
\begin{align}
    H_1 = \sum_{j=1}^{L-1} \left[-J a_{j}^{\dagger}a_{j+1}+\Delta a_{j}a_{j+1} + \text{H.c.}\right] -\mu \sum_{j=1}^{L} a_{j}^{\dagger}a_{j},
    \label{KC1}
\end{align}
where $a_j (a_j^{\dagger})$ is the spinless annihilation (creation) operator at the site $j$. The parameters $J$, $\Delta$, and $\mu$ correspond to the hopping amplitude, the $p$-wave pairing strength, and the chemical potential, respectively. The first Kitaev chain supports a non-trivial topological phase, that supports MZMs localized at its edge, when condition $|\mu|<2J$ is satisfied; otherwise, a trivial phase without edge modes. Similarly, the second Kitaev chain Hamiltonian $H_2$ is
\begin{align}
    H_{2} = \sum_{j=1}^{L-1} \left[-J b_{j}^{\dagger} b_{j+1} + \Delta e^{-i\phi} b_{j} b_{j+1} + \text{H.c.} \right] - \mu \sum_{j=1}^{L} b_{j}^{\dagger} b_{j},
    \label{KC2}
\end{align}
where $b_j (b_j^\dagger)$ denotes the annihilation (creation) operator at the site $j$ of the second Kitaev chain, and the phase factor $e^{-i\phi}$ characterizes the superconducting phase-difference with respect to the first Kitaev chain. In the absence of diagonal coupling between the chains, $H_2$ describes an independent topological superconducting wire that supports MZMs localized at its edges in the topological phase. Furthermore, the coupling between the two Kitaev chains is given as
\begin{align}
    H_{12}= \sum_{j=1}^{L-1} \left[ J_h a_{j}^{\dagger}b_{j+1} + J_h' b_{j}^{\dagger}a_{j+1} +  \text{H.c.} \right].
    \label{KC_coupled}
\end{align}
Here, $J_h$ and $J_h'$ represent the diagonal inter-chain coupling strength between the adjacent sites of the two chains. The term $J_h a_{j}^{\dagger}b_{j+1}$ denotes the coupling of fermions from the site $j$ of chain $a$ to the site $j+1$ of chain $b$, while the term $J_h' b_{j}^{\dagger}a_{j+1}$ represents the hopping of fermions from the site $j$ of chain $b$ to the site $j+1$ of chain $a$. The Hermitian conjugate (H.c.) terms ensure that the Hamiltonian remains Hermitian. Both diagonal couplings $J_h$ and $J_h'$ can be tuned to explore different topological regimes. A schematic illustration of the zigzag Kitaev chain Hamiltonian [Eq.~\eqref{H_ZKC1}] is shown in Fig.~\ref{two_kitaev_parallel_1}. Moreover, we assume that all the model parameters $J, \Delta, J_h$, and $J_h'$ are real.
\subsection{Energy Spectrum}\label{ES_ZKC}
\begin{figure}[htbp]
\includegraphics[width=0.49\linewidth,height=0.48\linewidth]{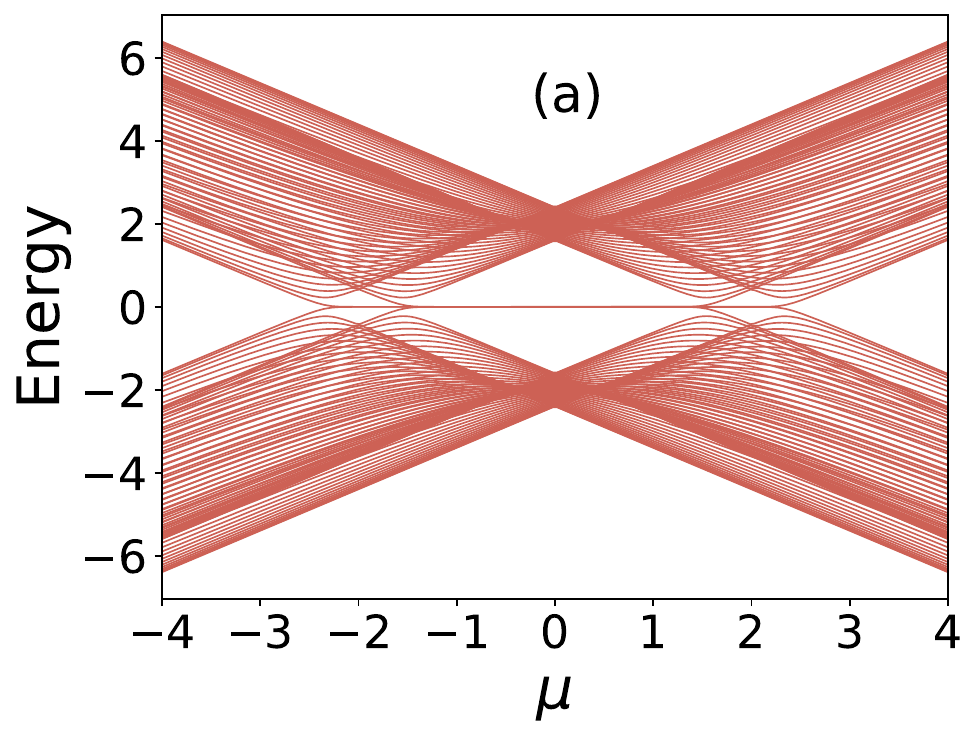}
\includegraphics[width=0.49\linewidth,height=0.48\linewidth]{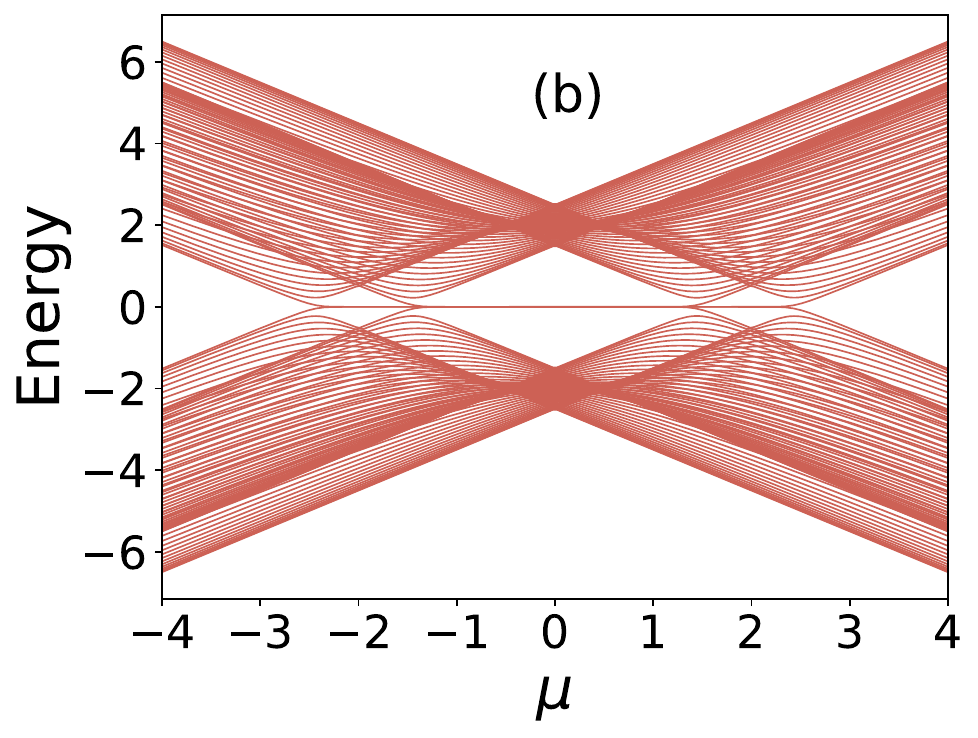}
\caption{Energy spectrum of the zigzag Kitaev chain with increasing $\mu$ for fixed $J=\Delta=1.0$ and $\phi=0$ chain lengths $N= 2L = 80$. Panels (a) and (b) correspond to asymmetric $(J_h, J_h')=(0.15,0.25)$ and symmetric $(J_h, J_h')=(0.25,0.25)$ inter-chain couplings, respectively.} 
\label{EG_ZKC_ES}
\end{figure}

In Fig.~\ref{EG_ZKC_ES}, the energy spectrum of the zigzag Kitaev chain with increasing the chemical potential \(\mu\) for fixed hopping amplitude and pairing strength \(J=\Delta=1.0\) and different diagonal inter-chain couplings \(J_h\) and \(J_h'\) in OBC. For suppressed coupling strength $J_h = J_h^{'}=0.0$, the energy spectrum closely resembles that of an isolated Kitaev chain, which exhibits only two transition points $\mu = \pm 2J$ associated with topological phase transitions. The Four MZMs appear in the range $-2J<\mu<2J$.
\begin{figure}[htb]
	\includegraphics[width=0.49\linewidth,height=0.48\linewidth]{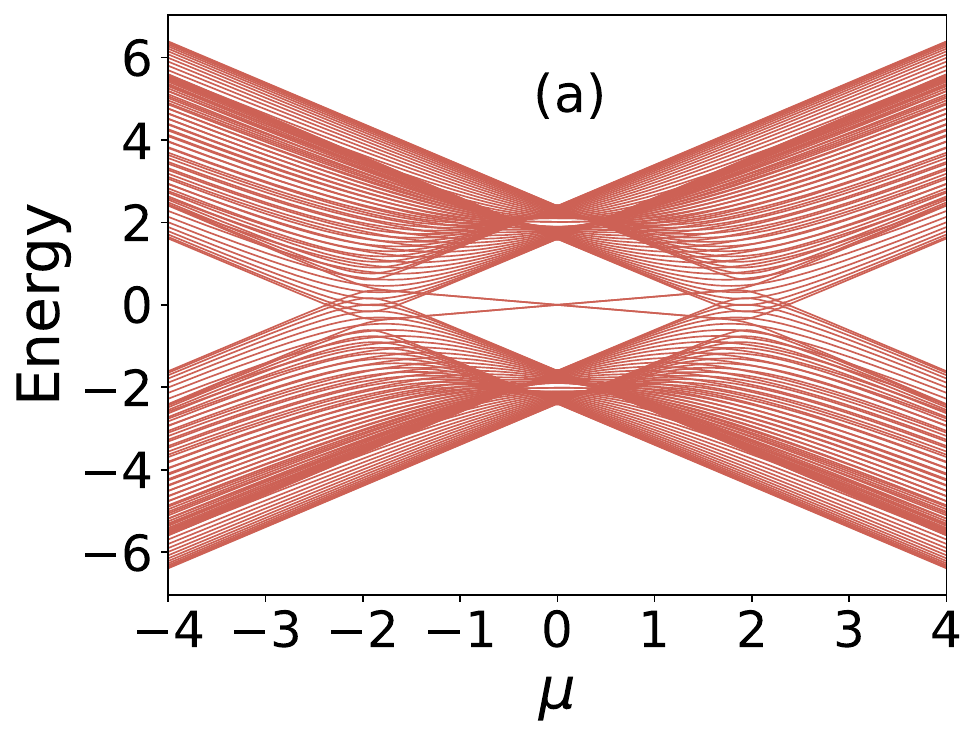}
	\includegraphics[width=0.49\linewidth,height=0.48\linewidth]{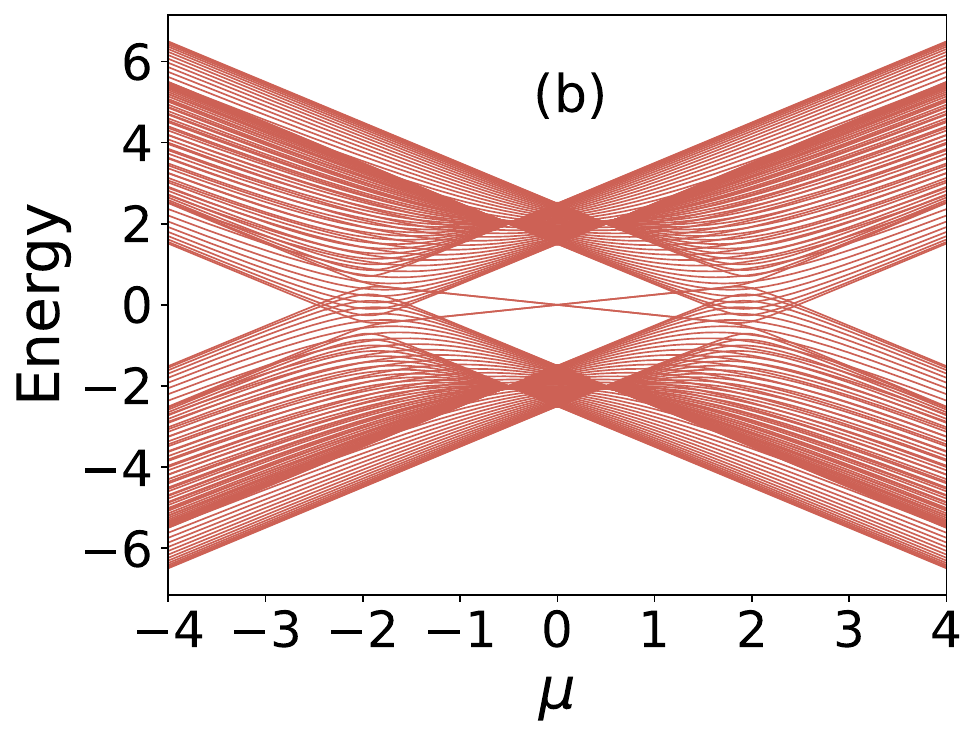}
	\caption{Energy spectrum of the zigzag Kitaev chain with increasing $\mu$ for fixed $J=\Delta=1.0$ and $\phi=\pi$ chain lengths $N= 2L = 80$. Panels (a) and (b) correspond to asymmetric $(J_h, J_h')=(0.15,0.25)$ and symmetric $(J_h, J_h')=(0.25,0.25)$ inter-chain couplings, respectively.} 
	\label{EG_ZKC_ES_phi_pi}
\end{figure}

In the zigzag Kitaev chain, such as \(J_h\) or \(J_h'\), finite and additional transition regions appear, shown in the energy spectrum [Fig.~\ref{EG_ZKC_ES}]. The energy spectrum [Fig.~\ref{EG_ZKC_ES}(a)] corresponds to different coupling strengths $(J_h \neq J_h^{'})$, and the energy spectrum [Fig.~\ref{EG_ZKC_ES}(b)] for the same coupling strength $(J_h = J_h^{'})$.  These regions arise from $J_h$ and  $J_h^{'}$, which enhance hybridization between MZMs residing on different Kitaev chains, and the hybridization develops multiple transitions within a finite range of \(\mu = |\pm 2J\pm (J_h+J_h^{'})|\), revealing the emergence of topological phases. The energy spectrum [Fig.~\ref{EG_ZKC_ES}] is symmetric under transformation $\mu \leftrightarrow -\mu$, reflecting particle-hole symmetry.

The energy spectrum for a superconducting pairing phase difference $\phi=\pi$ is shown in Fig.~\ref{EG_ZKC_ES_phi_pi}. In this case, the energy spectrum shows a lifting of MZMs degeneracies compared to the case $\phi=0$. For $\phi=0$, the system supports four degenerate MZMs in region I, while in region II the degeneracy is reduced to two MZMs, as discussed in Fig.~\ref{EG_ZKC_ES}. In contrast, for $\phi=\pi$, region I no longer supports four degenerate MZMs, and in region II only two MZMs persist, indicating a partial lifting of degeneracy, as shown in Fig.~\ref{EG_ZKC_ES_phi_pi}. The energy spectrum in Fig.~\ref{EG_ZKC_ES_phi_pi}(a) corresponds to the case of different inter-chain coupling strengths, while Fig.~\ref{EG_ZKC_ES_phi_pi}(b) represents the case with equal inter-chain coupling strengths.  Furthermore, the energy spectrum exhibits multiple band crossing points within the chemical potential $|\mu| = (2J - J_h + J_h^{'})$ to $(2J + J_h + J_h^{'})$. These band crossings correspond to gap closings at transition points and show topological phase transitions. Consequently, within the transition points, the presence of multiple crossings increases the degeneracies of the system. Moreover, from the energy spectrum [Fig.~\ref{EG_ZKC_ES}], we identify three distinct parameter regions characterized by different numbers of MZM.
\begin{figure}[htb]
	\includegraphics[width=0.49\linewidth,height=0.48\linewidth]{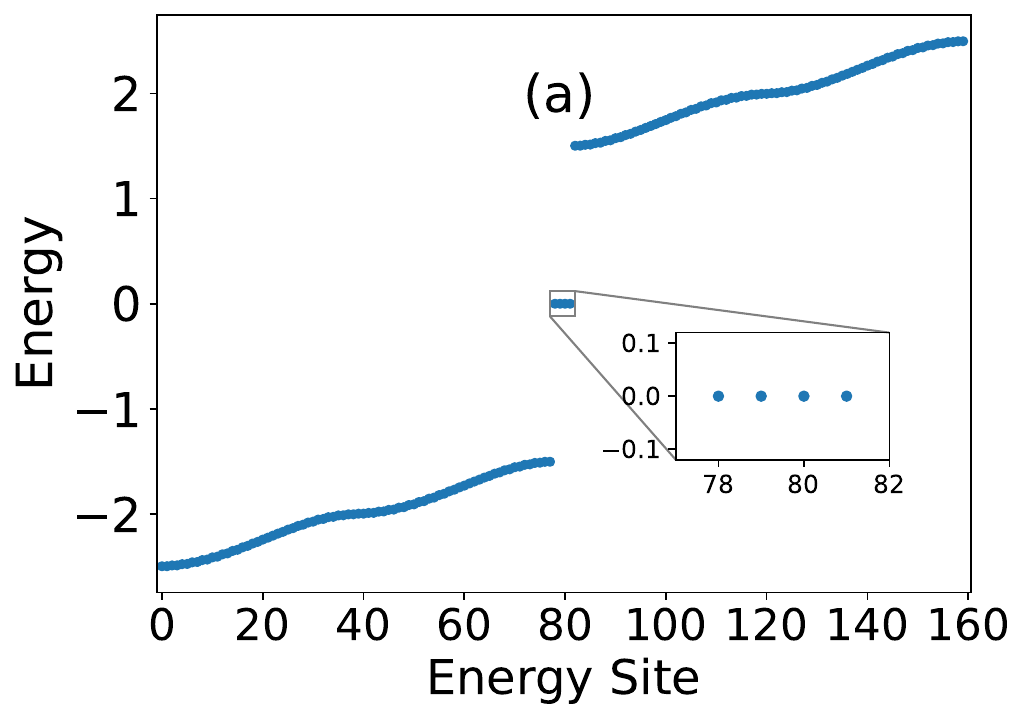}
	\includegraphics[width=0.49\linewidth,height=0.48\linewidth]{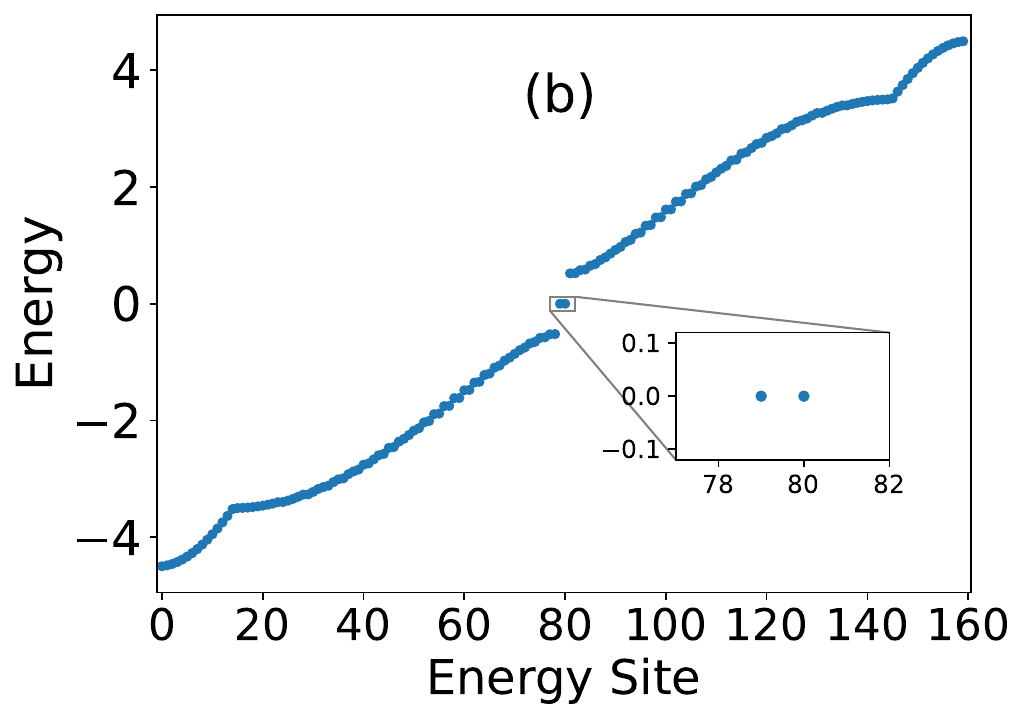}
	\caption{Eigenvalue spectrum as a function of the site index for the zigzag Kitaev chain with $J_h=J_h^{'}=0.25$, $J=\Delta=1.0$ chain lengths $N= 2L = 80$. Panels (a) and (b) show $\mu=0.0$ and $\mu=2.0$, respectively. The system hosts four MZMs in (a) and two MZMs in (b), as highlighted in the insets.}
	\label{EG_ZKC_EV}
\end{figure}

\textit{Region I:} For $-(2J - J_h -J_h') < \mu < (2J - J_h - J_h')$, the region supports four MZMs, as illustrated in Fig.~\ref{EG_ZKC_EV}(a). This regime corresponds to a higher-order topological phase, arising from the simultaneous presence of topological characters.

\textit{Region II:} For $-(2J + J_h + J_h') < \mu < -(2J - J_h - J_h')$ and $(2J - J_h - J_h') < \mu < (2J + J_h + J_h')$, both regions host two MZMs, as shown in Fig.~\ref{EG_ZKC_EV}(b). In this regime, only one sector remains topological, while the other becomes trivial. Both Region I and Region II correspond to topological phases, distinguished by different MZM degeneracies.

\textit{Region III:} For $|\mu| > (2J + J_h + J_h')$, the system does not have edge modes and is in a trivial phase.

The phase transition points are determined by the energy spectrum,
\begin{align}
\mu = \pm 2J \pm (J_h + J_h').
\label{transition_points}
\end{align}
These transition points are explicitly confirmed by the energy spectrum in Fig.~\ref{EG_ZKC_ES}, where the quasiparticle gap closes and reopens, signaling topological phase transitions with different numbers of MZMs. The resulting phase diagram from the energy spectrum is shown later. A detailed analytical characterization of the phase boundaries in momentum space and the associated topological invariants is provided in the next section.
\begin{figure}
\centering
\includegraphics[width=0.47\linewidth,height=0.40\linewidth]{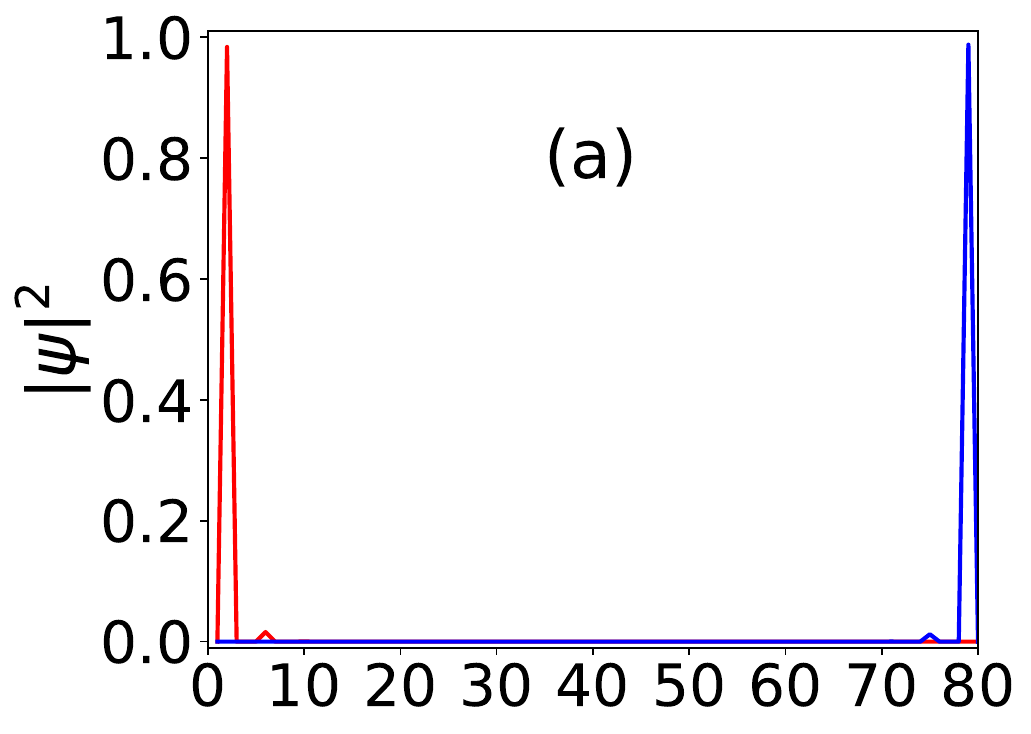}
\includegraphics[width=0.47\linewidth,height=0.40\linewidth]{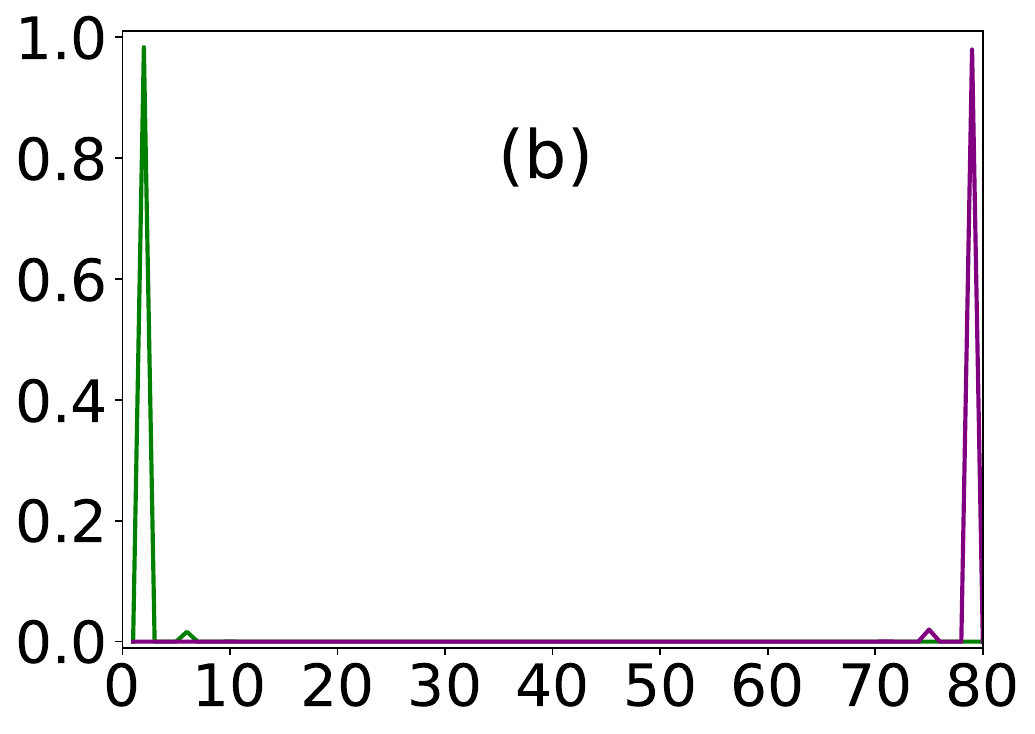}
\includegraphics[width=0.48\linewidth,height=0.45\linewidth]{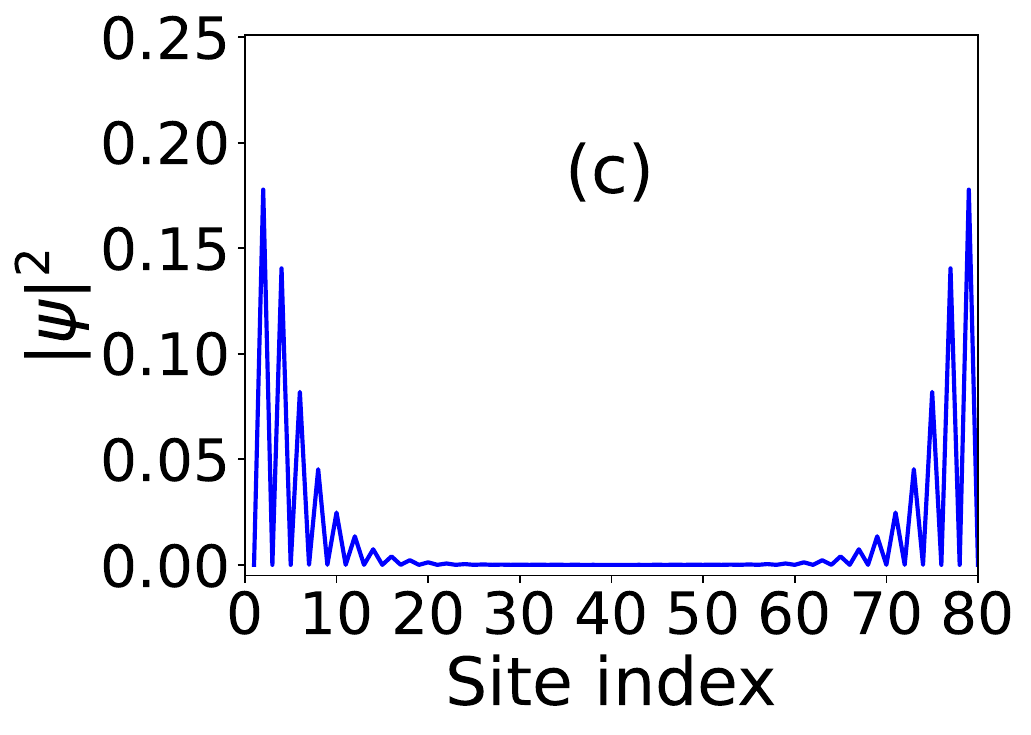}
\includegraphics[width=0.48\linewidth,height=0.45\linewidth]{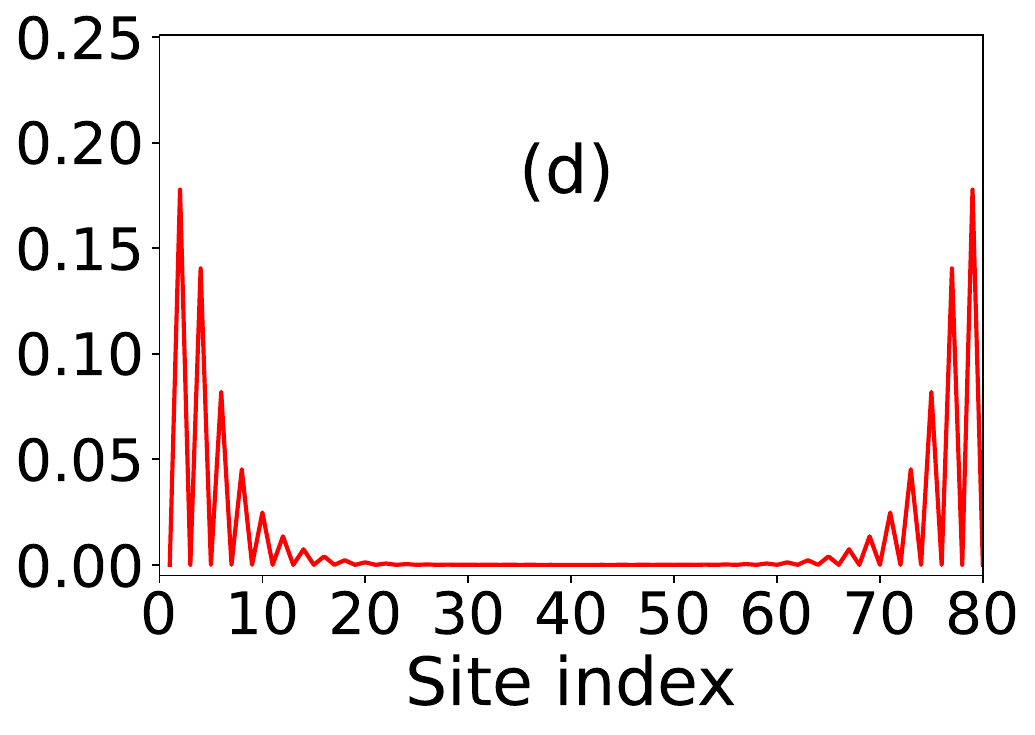}
\caption{The distribution of four MZMs in region I. For the parameters $J = \Delta = 1.0$ and diagonal coupling strength $J_h =J_h^{'} = 0.25$ and chemical potential $\mu$ (a,b)  0.0 and (c,d) 2.0.}
\label{EG_ZKC_MZM_Jh_0.25}
\end{figure}
\begin{figure}
\centering
\includegraphics[width=0.6\linewidth,height=0.5\linewidth]{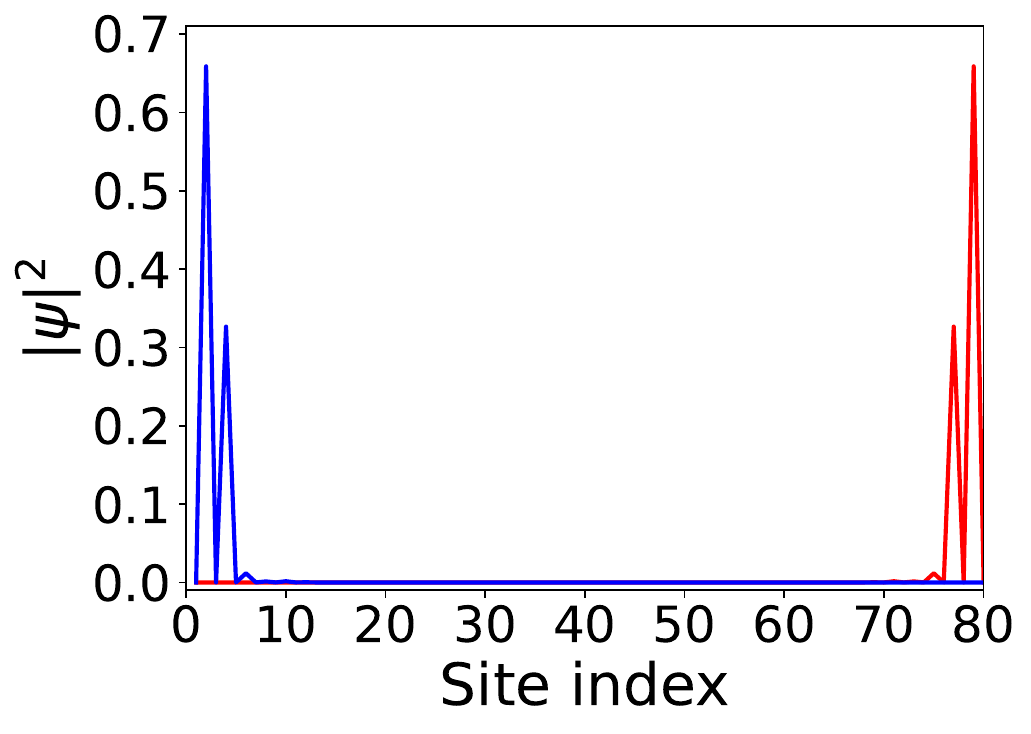}
\caption{The distribution of two MZMs. For the parameters $J = \Delta = 1.0$ and diagonal coupling strength $J_h = 0.25, J_h^{'} = 1.5$ and chemical potential $\mu = 2.0$.}
\label{EG_ZKC_EV_Jh_1.5}
\end{figure}
\subsection{Majorana zero modes and its distribution}\label{EV_ZKC}
To further examine the topological phases and the number of MZMs of the zigzag Kitaev chain for $\phi=0$, we analyze the energy eigenvalue spectrum for fixed $\mu$, according to the spectrum shown [Fig.~\ref{EG_ZKC_ES}]. In Fig.~\ref{EG_ZKC_EV}, we plot the eigenvalues with respect to the site index for \(J_h=J'_h=0.25\) and two representative values of $\mu$ that are zero in the topological region I and two in the topological region II. In region I, see Fig.~\ref{EG_ZKC_EV}(a) for $\mu=0$, the spectrum exhibits four zero-energy that are well separated from the bulk. These states correspond to MZMs of the zigzag chain. The distribution of these MZMs is shown in figures~\ref{EG_ZKC_MZM_Jh_0.25}(a) and~\ref{EG_ZKC_MZM_Jh_0.25}(b).  We observe that as the chemical potential increases (see also in Fig.~\ref{EG_ZKC_EV}(b) for $\mu=2.0$), the bulk gap is decreased and the zero energy states are reduced from four to two as shown in the distribution Figs.~\ref{EG_ZKC_MZM_Jh_0.25}(c)-(d). Furthermore, if the inter-chain coupling is increased, i.e., $J_h= 0.25$ and $J_h^{'} = 1.5$, the two MZMs are presented at the edges of the zigzag chain; see Fig.~\ref{EG_ZKC_EV_Jh_1.5}.  In this transitional regime, the MZMs begin to hybridize with bulk excitations, leading to a visible splitting in the MZMs states.  The presence of a finite bulk gap protects these MZMs against hybridization, confirming their existence and robustness of MZMs in topological regimes.
\subsubsection*{Majorana qubit encoding: tetron architectures}
The region I, supporting four MZMs in the zigzag Kitaev chain, can be directly related to a minimal MZM based qubit encoding. In analogy to tetron architectures \cite{qx36-4rv1}, the four MZMs of the system define a nonlocal fermionic Hilbert space with fixed total parity. Within this subspace, effective Pauli operators can be constructed from bilinear Majorana operators that act on the encoded qubit states. In our zigzag model, the diagonal inter-chain couplings and chemical potential play a role analogous to tunable hybridization between MZMs, thereby controlling the splitting and stability of the qubit subspace.
\subsection{Bulk properties and dispersion relation}\label{BP_DR_Two_ZKC}
\begin{figure}[t!]
	\includegraphics[width=1.0\linewidth,height=0.37\linewidth]{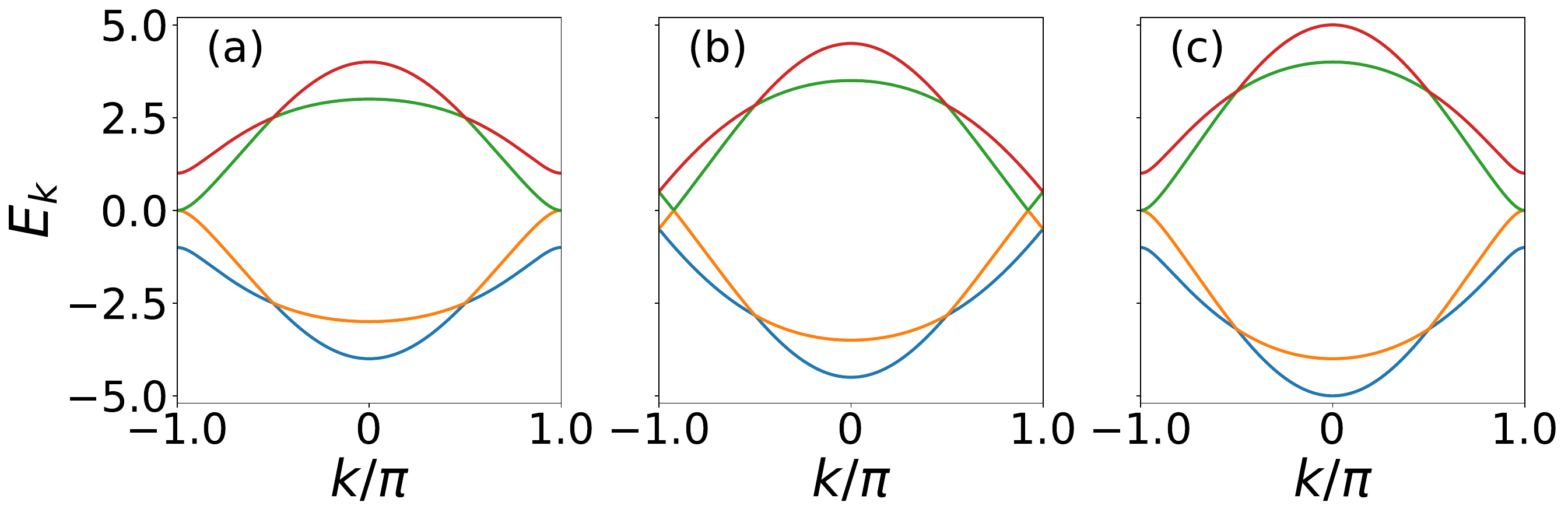}
	\includegraphics[width=1.0\linewidth,height=0.37\linewidth]{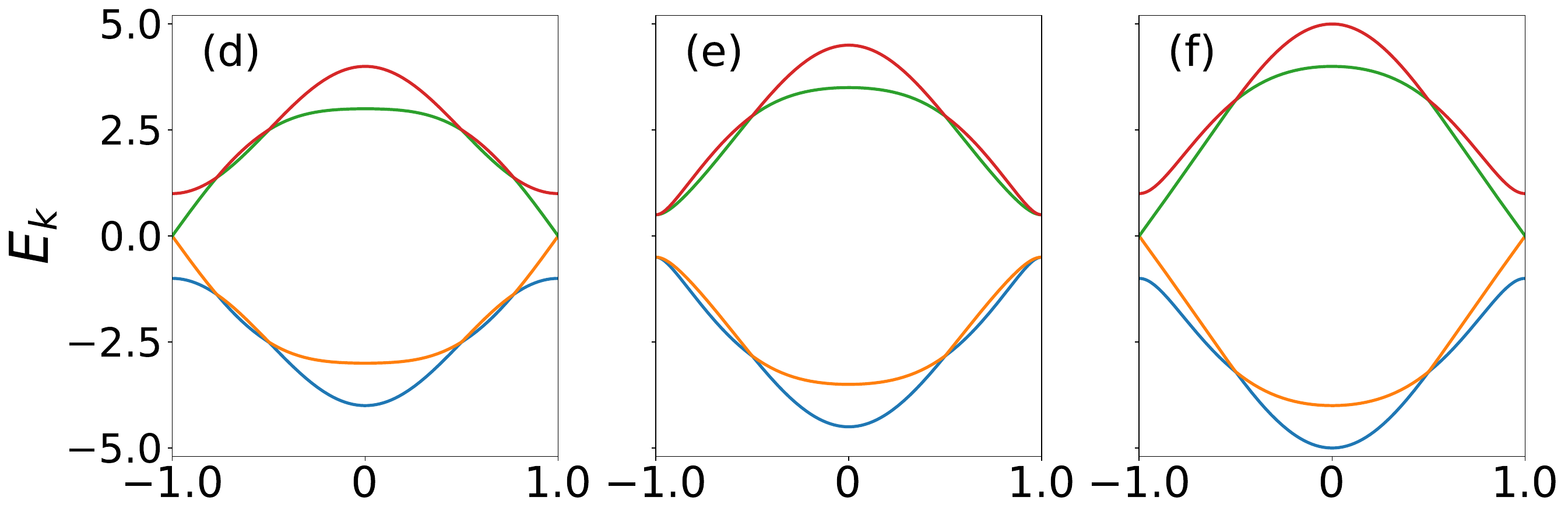}
	\caption{Dispersion relation of the zigzag Kitaev chain from Eq.~\ref{Eq:EnergySpectrum}.  The energy spectrum is plotted for $J = \Delta = 1.0$ and \(J_h = J_h^{'} = 0.25\) for different $\mu$ values: (a) $1.5$, (b) $2.0$, and (c) $2.5$ for phase difference $\phi=\pi$; however, for $\mu$ (d) $1.5$, (e) $2.0$, and (f) $2.5$ for phase difference $\phi=0$.}
	\label{EG_ZKC_DR}
\end{figure}

We write the zigzag Hamiltonian [Eq.~(\ref{H_ZKC1})] in momentum space using the Fourier transform. We also introduce the Nambu spinor $\Psi_k^\dagger = (a_k^\dagger, b_k^\dagger, a_{-k}, b_{-k})$ in the particle-hole basis to write it in the Bogoliubov-de Gennes (BdG) form
\begin{equation}
H = \frac{1}{2}\sum_k \Psi_k^\dagger H_{\mathrm{BdG}}(k)\Psi_k, 
\end{equation}
where
\begin{equation}
H_{\mathrm{BdG}}(k)=
\begin{pmatrix}
\xi_k & J_k & \Delta_k & 0 \\
J_k^* & \xi_k & 0 & \Delta_k e^{-i\phi} \\
-\Delta_k & 0 & -\xi_k & -J_k \\
0 & -\Delta_k e^{i\phi} & -J_k^* & -\xi_k
\end{pmatrix}
\label{BdG_ZKC},
\end{equation}
with $J_k = J_h e^{ik} + J_h' e^{-ik}$, $\xi_k = -\mu - 2J\cos k$, and $\Delta_k = 2i\Delta\sin k$. Diagonalizing the Hamiltonian in Eq.~(\ref{BdG_ZKC}) yields four eigenvalues
\begin{align}
	E_k =  \pm\sqrt{ \xi_k^2 + |J_k|^2 + |\Delta_k|^2\pm2 |J_k|\sqrt{ \xi_k^2  + |\Delta_k|^2  \sin^2(\frac{\phi}{2})}}.
	\label{Eq:EnergySpectrum}
\end{align}
The band structure of the zigzag Kitaev chain clearly consists of four bands originating from bonding and anti-bonding states in the particle and hole channels. Here, the outer $\pm$ corresponds to particle-hole symmetry, while the inner $\pm$ distinguishes the bonding and anti-bonding quasiparticle bands arising from the inter-chain coupling.  The superconducting phase difference $\phi$ enters the energy spectrum through the term $\sin^2(\phi/2)$, reflecting the interference between the pairing amplitudes of the two chains. When $\phi=\pi$, the quasi-particle dispersion relation in Eq.~(\ref{Eq:EnergySpectrum}) becomes
\begin{equation}
E_k= \sqrt{\xi_k^2+|\Delta_k|^2} \pm |J_k|,
\label{DR_phase_pi}
\end{equation}
which describes two effectively decoupled Kitaev chains whose energy dispersions are shifted by $\pm |J_k|$, see Figs.~\ref{EG_ZKC_DR}(a)-(c).
However, the quasi-particle dispersion relation in Eq.~(\ref{Eq:EnergySpectrum}) simplifies to
\begin{equation}
E_k= \pm\sqrt{(\xi_k \pm |J_k|)^2 + |\Delta_k|^2 },
\label{DR_phase_zero}
\end{equation}
when $\phi=0$. The resulting band structure is illustrated in Figs.~\ref{EG_ZKC_DR}(d)-(f) at the transition points $\mu = 1.5$ and $2.5$. The bands become gapless at $k=0$ and $\pi$, indicating bulk gap closings at these momenta. For diagonal coupling strengths $J_h = J_h^{'} = 0.25$, both values of the chemical potential satisfy the topological phase transition points.

The bulk phases of the system are determined by the closing and reopening of the quasiparticle excitation gap. The gap closes when both conditions $|\Delta_k|=0$ and $\xi_k \pm |J_k|=0$ are simultaneously satisfied. The pairing amplitude $|\Delta_k|$ vanishes at the high-symmetry momenta $k=0, \pi$. Collecting all possibilities, the phase boundaries can be found as
\begin{equation}
\mu = \pm 2J \pm |J_h + J_h'| .
\label{T_Phase_point}
\end{equation}
For equal inter-chain coupling strengths, i.e. $J_h^{'} = J_h$, the coupling term simplifies to $J_k = 2J_h \cos k$. Using this form, we show the dispersion in Figs.~\ref{EG_ZKC_DR}(d) and~\ref{EG_ZKC_DR}(e) for $\mu = 1.5$ and Fig.~\ref{EG_ZKC_DR}(f) for $\mu = 2.5$. We observe that the spectrum remains fully gapped except when the chemical potential is tuned to the critical values $\mu = \pm 2\left(J \pm J_h\right)$, where gap closings occur, corresponding to topological phase transitions. These transition points separate the distinct gapped phases, corresponding to the parameter regimes $|\mu|>2\left(J + J_h\right)$ as a trivial phase, $2\left(J + J_h\right) > |\mu| > 2\left(J - J_h\right)$, and $|\mu| < 2\left(J - J_h\right)$ as topological phases with two and four MZMs. These boundaries are the reduced form of the transition points given in Eq.~(\ref{T_Phase_point}).
\subsection{Symmetry and Winding Number}\label{WN_ZKC}
To characterize the topological phases, we calculate the integer-valued winding number ($\nu$) \cite{Nehra_2020, PhysRevB.101.214507, PhysRevB.103.224208} in one-dimensional systems, which depends on the number of effective edge modes and quantized winding number values such as $\nu=0, 1$ and $2$. The $\nu=0$ corresponds to a trivial phase, while $\nu=1$ and $2$ emerge as the topological phase with two and four MZMs. The transitions between distinct topological phases occur at transition points where the bulk quasienergy gap closes. In particular, recent studies have reported localized edge modes, motivating a detailed analysis of the winding number in the vicinity of transition points to establish the correspondence with boundary excitations~\cite{PhysRevLett.120.057001, PhysRevLett.111.173004, PhysRevX.7.041048}.
\begin{figure}[t!]
	\includegraphics[width=0.88\linewidth, height=0.6\linewidth]{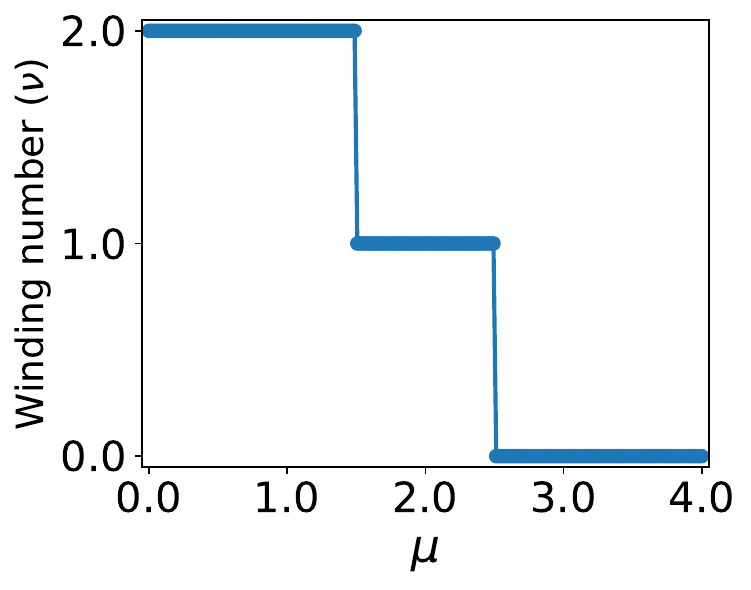}
	\caption{Winding number $\nu$ as a function of chemical potential $\mu$ for $J=\Delta=1$ and symmetric diagonal coupling $J_h=J'_h=0.25$. The stepwise changes in $\nu$ occur at $|\mu|=2(J\pm J_h)$, indicating successive topological phase transitions between phases with $\nu=2,1$  and $0$.}
	\label{EG_ZKC_WN}
\end{figure}


The zigzag Kitaev chain has an integer winding number. Depending on the system parameters, the model exhibits phases with $\nu=2, 1$ and $0$. The highest winding number $\nu=2$ corresponds to the presence of two MZM per edge, while $\nu=1$ indicates a MZM per edge and $\nu=0$ denotes the absence of edge modes. For Winding number calculation, the BdG Hamiltonian in Eq.~(\ref{BdG_ZKC}) with superconducting pairing phase difference zero ($\phi=0$) can be written in block form,
\begin{align}
H(k)= \begin{pmatrix}
h(k) & \Delta(k) \\
\Delta^\dagger(k) & -h(k)
\end{pmatrix},
\label{Block_H}
\end{align}
with
\begin{align}
h(k)&=\begin{pmatrix}\xi_k & J_k \\ J_k^* & \xi_k\end{pmatrix} \quad\mathrm{and}\quad
\Delta(k)=\begin{pmatrix}\Delta_k & 0 \\ 0 & \Delta_k\end{pmatrix}.
\end{align}
We now analyze the discrete symmetries of the Hamiltonian in Eq.~(\ref{Block_H}), which satisfied particle-hole $(C)$, time-reversal $(T)$, and chiral (S) symmetry:
\begin{align}
&\mathcal{C} H(k)\mathcal{C}^{-1}=-H(-k), \quad \mathcal{C}=(\tau_x \otimes I_2) K, \\
&\mathcal{T}H(k)\mathcal{T}^{-1}=H(-k),\quad \mathcal{T}=I_4 K,
\end{align}
where $K$ denotes complex conjugation, $\tau_{i=x,y,z}$ is the Pauli matrices acting in particle-hole space, and $I$ is the identity matrix. In the chiral basis, the BdG Hamiltonian [Eq.~(\ref{BdG_ZKC})] can be brought into an off-diagonal form,
\begin{align}
H_{\mathrm{off}}(k)= U H(k) U^\dagger = 
\begin{pmatrix}
0 & Q(k) \\
Q^\dagger(k) & 0
\end{pmatrix}
,\end{align}
where $U$ denotes the basis change unitary transformation to the chiral basis and the off-diagonal block
\begin{equation}
Q(k)=
\begin{pmatrix}
\xi_k + \Delta_k & J_k \\
J_k^{*} & \xi_k + \Delta_k
\end{pmatrix}.
\end{equation}
\begin{figure}[t!]
	\includegraphics[width=0.9\linewidth, height=0.6\linewidth]{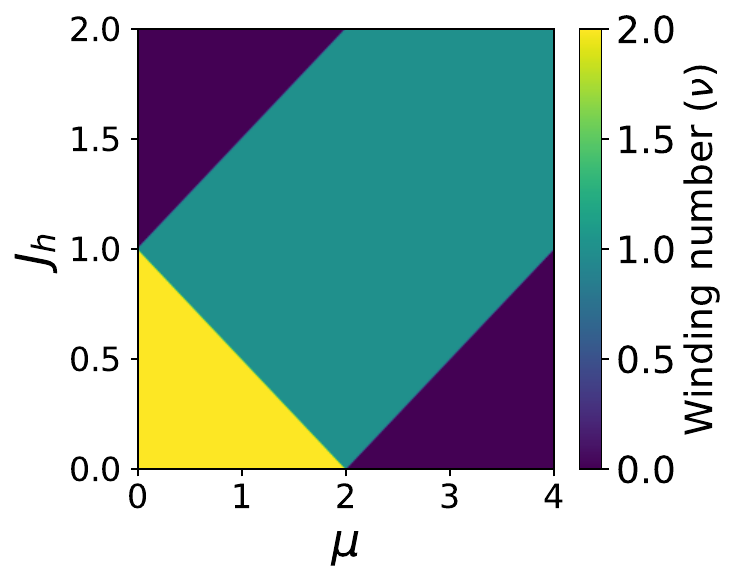}
	\caption{Phase diagram in ($\mu, J_h$) plane obtained by calculating the winding number $\nu$ for the inter-chain coupling $J'_h = J_h$ with $J=\Delta=1$. Clearly, it shows three topologically distinct regions with $\nu=0, 1,$ and $2$, corresponding to the trivial phase and phases supporting two and four MZMs per edge, respectively. The phase boundaries follow the conditions $|\mu|=2(J\pm J_h)$ where the bulk-gap closes.}
	\label{EG_ZKC_WN1}
\end{figure}
\begin{figure*}[t!]
	\includegraphics[width=0.32\linewidth,height=0.27\linewidth]{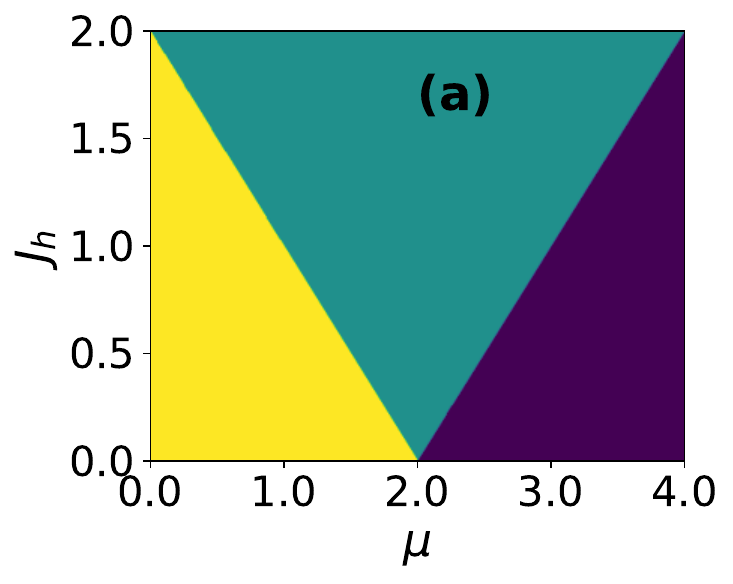}
	\includegraphics[width=0.30\linewidth,height=0.27\linewidth]{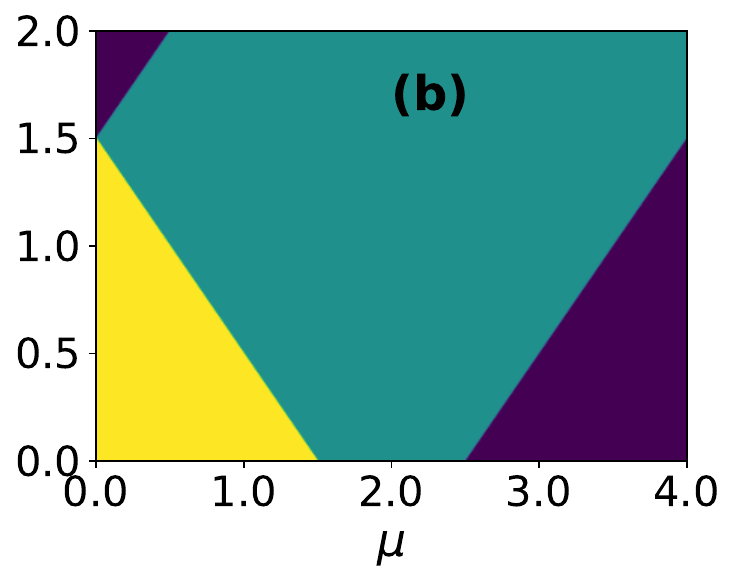}
	\includegraphics[width=0.35\linewidth,height=0.27\linewidth]{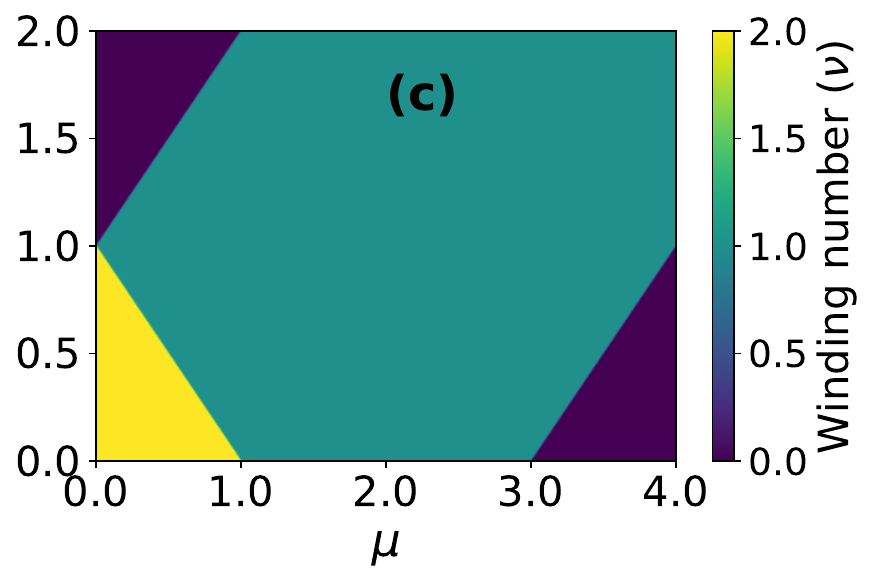}
	\caption{Phase diagram in ($\mu, J_h$) plane obtained from the winding number $\nu$ for fixed parameters $J=\Delta=1$ and different values of $J_h'$. (a) $J'_h = 0$, (b) $J'_h = 0.5$,  and (c) $J'_h = 1$. The color scale indicates the winding number. The distinct colored regions correspond to different topological phases with $\nu=0,1,2$, representing the trivial phase, the topological phase with two MZMs, and four MZMs of the zigzag chain, respectively.}
	\label{EG_ZKC_WN2}
\end{figure*}
This off-diagonal representation is particularly convenient because the topology of the system is fully encoded in the matrix $Q(k)$. For chiral systems, the winding number $\nu$ is defined as
\begin{equation}
\nu = -\frac{1}{2\pi i} \int_{-\pi}^{\pi} \partial_k \ln \det Q(k)\ dk.
\end{equation}
The winding number $\nu$ provides a clear signature of the multiple MZMs in the topological phases. In Fig.~\ref{EG_ZKC_WN}, we present $\nu$ as a function of $\mu$, for the parameters $J=\Delta=1$ and $J_h=J_h^{'} = 0.25$. The integer values of $\nu$ correspond directly to the number of MZMs localized at the edges, reflecting the bulk-boundary correspondence. For symmetric diagonal couplings $J_h=J_h'$, the winding number is obtained in the parameters regime as
\begin{align}
\nu =
\begin{cases}
0, & |\mu| > 2J + 2J_h, \\[2pt]
1, & 2J - 2J_h < |\mu| < 2J + 2J_h, \\[2pt]
2, & |\mu| < 2J - 2J_h,
\end{cases}
\label{WN_critical_sym}
\end{align}
which predicts two topological phase boundaries at $|\mu|=2(J\pm J_h)$. These transition points coincide with bulk gap closings in the quasi-dispersion spectrum and energy spectra. For $J=\Delta=1$ and $J_h=J_h'=0.25$, the transitions occur in $\mu=1.5$ and $\mu=2.5$, separating phases with $\nu=2,1$ and $0$, corresponding to four, two MZMs, and without edge modes, respectively. The numerical results shown in Fig.~\ref{EG_ZKC_WN} are in excellent agreement with the analytical prediction, exhibiting stepwise changes in $\nu$ that reflect the closing and reopening of the bulk gap across the topological phase transitions.

The topological phase diagram for the symmetric inter-chain coupling strength $J_h = J_h'$ is shown in Fig.~\ref{EG_ZKC_WN1}, where the winding number $\nu$ is plotted as a function of the chemical potential and the diagonal coupling strengths for fixed parameters $J=\Delta=1$. The diagram exhibits three distinct regions characterized by $\nu=0, 1,$ and $2$, corresponding to the trivial phase and topological phases with two and four MZMs, respectively. The phase boundaries form symmetric linear contours in the $(\mu, J_h)$ plane and are determined by the analytical gap-closing conditions $|\mu|=2(J\pm J_h)$. In particular, increasing $J_h$ expands the topological regions with higher winding number, with multiple MZMs. The numerical results are in excellent agreement with the analytical predictions, confirming the robustness of the multichannel topological phases in the symmetric zigzag Kitaev chain.

The topological phase diagram with an asymmetric inter-chain coupling strength is shown in Fig.~\ref{EG_ZKC_WN2}, where the winding number $\nu$ is plotted as a function of the chemical potential $\mu$ and the coupling strength $J_h$ for fixed parameters $J=\Delta=1$ and different values of $J_h'$. Panels (a)–(c) correspond to $J_h'=0$, $0.5$, and $1$, respectively. The phase diagrams exhibit three distinct regions characterized by $\nu=0, 1,$ and $2$, corresponding to the trivial phase and topological phases supporting two and four MZMs. As $J_h'$ increases, the phase boundaries shift and become asymmetric, leading to a deformation of the topological regions in the $(\mu, J_h)$ plane. In particular, the region supporting the higher winding number ($\nu=2$) is progressively modified, reflecting the imbalance between the two inter-chain couplings. These results demonstrate that asymmetric coupling provides an additional tuning parameter to control the emergence and stability of multiple MZMs topological phases in the zigzag Kitaev chain. In contrast to the symmetric case, where the phase boundaries are symmetric in the $(\mu, J_h)$ plane, the introduction of asymmetric inter-chain coupling ($J_h \neq J_h'$) breaks this symmetry and leads to a deformation and shift of the topological regions, thereby modifying the stability and extent of phases hosting multiple MZMs. From the winding number, we examine the transition points in the parameters regime as
\begin{align}
\nu =
\begin{cases}
0, & |\mu| > 2J + J_h + J_h', \\[2pt]
1, &2J - (J_h + J_h') < |\mu| < 2J + J_h + J_h', \\[2pt]
2, & |\mu| < 2J - (J_h + J_h'),
\end{cases}
\label{WN_critical_asym}
\end{align}
where the three regimes correspond to the trivial phase and topological phases with two and four MZMs, respectively. These results are consistent with the phase transition points given by Eq.~(\ref{T_Phase_point}), $\mu = \pm 2J \pm |J_h + J_h'|$ for asymmetric diagonal inter-chain coupling and by Eq.~(\ref{transition_points}), confirming that the changes in winding number $\nu$ coincide with the bulk-gap closing conditions obtained from the dispersion relation.
\section{Phase Diagram}\label{PD_ZKC}
\begin{figure}
\begin{tikzpicture}[xscale=1.05, yscale=3]
\fill[gray!30] (-3.55,-0.15) rectangle (-2.5,0.15);
\fill[gray!30] (2.5,-0.15) rectangle (3.55,0.15);
\fill[orange!40] (-2.5,-0.15) rectangle (-1.5,0.15);
\fill[orange!40] (1.5,-0.15) rectangle (2.5,0.15);
\fill[cyan!50] (-1.5,-0.15) rectangle (1.5,0.15);
\draw[<->, line width=1.8pt] (-3.55, 0) -- (3.55, 0)
node[right, blue] {\large $\mu/J$};
\draw[line width=1.4pt] (-1.5, -0.15) -- (-1.5, 0.15);
\draw[line width=1.4pt] (-2.5, -0.15) -- (-2.5, 0.15);
\draw[line width=1.4pt] (1.5, -0.15) -- (1.5, 0.15);
\draw[line width=1.4pt] (2.5, -0.15) -- (2.5, 0.15);
\foreach \x in {-3,-2,-1,0,1,2,3} {
  \draw[line width=1pt] (\x,0.0) -- (\x,-0.04);
  \node[below] at (\x,0) {\large \x};
}
\end{tikzpicture}
\caption{Phase diagram along $\mu/J$ line for $J=\Delta=1.0, J_h = J_h^{'}=0.25$. The cyan region with $\nu=2$ corresponding to four MZMs, appear in the interval $-2J+2J_h < \mu < 2J-2J_h$, i.e.  for $-1.5 < \mu < 1.5$. The orange regions with $\nu=1$ supporting two MZMs in $-2J-2J_h < \mu < -2J+2J_h$ and $2J-2J_h < \mu < 2J+2J_h$, i.e. for $-2.5 < \mu < -1.5$ and $1.5 < \mu < 2.5$. However, for gray region $|\mu| > 2J+2J_h$, i.e. for $|\mu| > 2.5$, the system with $\nu=0$ is in topologically trivial phase.}
\label{Phase_diagram_ZKC}
\end{figure}
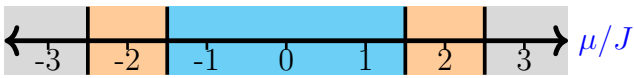

In this section, we finalize the topological phase diagram for the zigzag Kitaev chain with $\phi=0$ using the energy spectrum in Fig.~\ref{EG_ZKC_ES}, the dispersion relation in Fig.~\ref{EG_ZKC_DR}, and the winding number in Fig.~\ref{EG_ZKC_WN}. We identify three distinct parameter regions, each characterized by a different number of MZMs and the corresponding quantized winding number. The transition points occur at
\begin{align}
\mu = \pm 2J \pm (J_h + J_h').
\label{transition_points_1}
\end{align}
These transition points are explicitly confirmed by all observables where the quasiparticle gap closes and reopens at the corresponding chemical potentials, showing transitions between topological and trivial phases of the zigzag Kitaev chain. In Fig.~\ref{Phase_diagram_ZKC}, the phase diagram shows the non-trivial phases along the $\mu/J$ parameter values. These transition points obtained coincide exactly with the transition points extracted from the energy spectrum, confirming the consistency between the analytical analyses, such as quasi-dispersion and winding number. Between these phase boundaries, the system realizes different topological phases characterised by the winding numbers $\nu = 0, 1$ and $2$ which correspond to zero, two, or four MZMs localized at the edges of the zigzag chain. For the region $-2J+2J_h < \mu < 2J-2J_h$, the phases are topologically nontrivial, resulting in a phase with a winding number $\nu=2$ and four MZMs. The intervals $-2(J+J_h) < \mu < -2(J-J_h)$ and $2(J-J_h) < \mu < 2(J+J_h)$, resulting in a topological phase with \(\nu=1\), that supports two MZMs. Outside these regions for $|\mu| > 2(J+J_h)$, the system enters a topologically trivial phase with \(\nu=0\) and does not have edge modes. The resulting phase diagram is summarized in Fig.~\ref{Phase_diagram_ZKC}, illustrating the sequence of topological transitions driven by the relative strength of the chemical potential and the inter-chain couplings.
\section{Conclusion}
\label{Conclusion}

In this work, we presented a theoretical investigation of a zigzag Kitaev chain, composed of two Kitaev chains connected through asymmetric diagonal inter-chain couplings. Our analysis started with finding the energy spectra of the system. We derived the corresponding BdG Hamiltonian in momentum space and obtained the bulk quasiparticle dispersion relation. The analytical energy dispersion reveals that the diagonal inter-chain coupling hybridizes the two Kitaev chains into bonding and anti-bonding sectors, leading to modified bulk bands and tunable superconducting gaps. The topological properties of the system were investigated using the winding number, a topological invariant, and the corresponding spatial distribution of MZMs. By identifying the bulk gap-closing conditions, we established the phase boundaries separating topological and trivial phases. The winding number accurately characterizes these phases and is fully consistent with the bulk-boundary correspondence. In the topological phase, the energy spectrum exhibits robust MZMs, whereas there are no such edge modes in the trivial phase. We further demonstrated that the asymmetric diagonal couplings provide an additional degree of freedom to control the bulk bands and the stability of the topological phases. The competition between the chemical potential and diagonal coupling hybridization gives rise to a rich phase diagram, illustrating the versatility of the zigzag system. Overall, our analytical results, corroborated with the energy spectrum, demonstrate the consistency between bulk gap closings, winding number transitions, and edge-state degeneracies. 

Recently, extensive experimental and theoretical efforts have advanced the exploration of the identification of MZMs~\cite{nanolett_7b01728}. A variety of experimental platforms have been proposed to realize topological superconductors, including semiconductor-superconductor heterostructures, and trapped-ion systems \cite{PhysRevA.72.063407, PhysRevA.99.052342}. In particular, quantum-dot arrays provide exceptional tunability via gate-controlled potentials, enabling controlled realization of coupled and dimerized Kitaev chains~\cite{PhysRevB.100.205302}. Our analysis offers experimentally relevant criteria for identifying and controlling distinct topological phases. Beyond solid-state implementations, the present model can be easily extended to engineering platforms, including photonic \cite{Douglas_2015}, electric~\cite{Iizuka_2023}, and mechanical systems \cite{Allein_2023}. The tunable emergence of multiple MZM demonstrated in this work highlights zigzag Kitaev chain as promising candidates for studying controllable Majorana hybridization and for developing MZM based qubit platforms in quantum dot arrays, superconducting circuits, and hybrid qubit photon systems \cite{Kanungo2022, Dennis_2002, PhysRevApplied.18.054037, science_aar4005}.

We also outline possible future directions of our work. Our findings provide a unified analytical and numerical framework for understanding zigzag Kitaev chains and offer useful guidance for future directions of topological systems. This model may also serve as a platform for investigating the effects of disorder, long-range superconducting pairing, Floquet driving, and non-Hermitian perturbations, as well as for exploring a deeper understanding of such systems and experimentally relevant semiconductor-superconductor nanowires aimed at realizing robust topological quantum devices.
\section*{Acknowledgement}
Rajiv Kumar acknowledges  financial support from the Council of Scientific and Industrial Research (CSIR) fellowship [File No. 09/1217(15934)/2022-EMR-I], New Delhi, India 
\section*{DATA AVAILABILITY}
\label{DATA}
The data that support the findings of this article are available from the authors upon reasonable request.
\bibliography{ref}
\end{document}